\documentstyle[12pt]{article}
\newcommand{\labe}[1]{\label{#1}}
\begin{document}

\title{The role of zero modes for the infrared behavior of QCD}
\author{Dieter Stoll, Soh Takeuchi and Koichi Yazaki\\ Department of Physics,
University of Tokyo\\ Bunkyo--ku, Tokyo 113, Japan\thanks{email:
 soh@tkyvax.phys.s.u-tokyo.ac.jp;
stoll@tkyux.phys.s.u-tokyo.ac.jp\hspace{.5cm}
FAX: (81)(3)56849642}}
\date{  }
\maketitle
\begin{abstract}
We analyse the mechanism in which zero modes lead to an elimination of
fermionic color non--singlet states in 1+1 dimensions. Using a hamiltonian
lattice formulation we clarify the physical meaning of the zero modes but we
do not find support for speculations on the crucial importance of lower
dimensional fields (zero modes in 1+1 dimension) for the infrared behavior
of QCD in 2+1 or higher dimensions.
\end{abstract}

\section{Introduction}
Within the last years, formulations of the QCD hamiltonian in terms of gauge
invariant degrees of freedom have been given
\cite{Goldstone,Lenz}. In such approaches the hamiltonian contains
singular coupling terms which resemble in many cases centrifugal barriers,
depending, however, on fields rather than on quantum mechanical variables. In
particular in an axial gauge representation of QCD these singular coupling
terms were found to be associated with "lower dimensional fields" which are
zero modes with respect to one coordinate direction \cite{Lenz}. Due to this
particular interaction a calculation in 1+1 dimensional Yang-Mills theory in
the presence of static charges has shown that the infrared properties
of the model crucially depend on these zero modes \cite{Schreiber}. Similar
observations have also been made in a study of Yang--Mills theory coupled
to non--relativistically moving particles \cite{Gupta}. In view of these
findings, the question has been
raised as to whether these lower dimensional fields may play a crucial role
for understanding the infrared properties of QCD and in particular confinement
in 3+1 dimensions, as well \cite{Lenz,Schreiber}.

In this letter we intend to address this question in the framework of
the lattice hamiltonian formulation of SU(N) Yang--Mills theory in the presence
of static color sources in 1+1 dimensions. In contrast to the
continuum formulation used in the aforementioned approaches we have a simple
physical interpretation for the gauge invariant degrees of freedom in the
lattice formulation. This enables us to interpret the "lower dimensional
fields"
which become quantum mechanical variables in 1+1 dimension. Therefore an
understanding of the role of zero modes can be
gained which is general enough to allow us drawing conclusions about
higher dimensions.

\section{Hamiltonian lattice QCD in 1+1 dimensions}\labe{SHLQCD}
\setcounter{equation}{0}
The variables in the hamiltonian formalism on the lattice are the link
variables
$\hat{U}(l);\ l=1\dots M$, which are group elements of SU(N), and
corresponding "angular momentum" operators $\hat{J}^a_R(l)$ and
$\hat{J}^a_L(l)$ which generate right and left multiplication in the
corresponding representation of the SU(N) group. They satisfy
the commutation relations
\begin{eqnarray}
\left[ \hat J^a_L(l) , \hat U_{ij}(l^\prime) \right]= (-T^a\hat U )_{ij}(l)
\delta_{l,l^\prime} && \left[ \hat J^a_L(l), \hat J^b_L(l^\prime) \right]=
i f^{abc}\hat J^c_L(l) \delta_{l,l^\prime} \nonumber    \\
\left[ \hat J^a_R(l) , \hat U_{ij}(l^\prime) \right]
= (-\hat U T^a )_{ij}(l) \delta_{l,l^\prime}
&&  \left[ \hat J^a_R(l), \hat J^b_R(l^\prime) \right]= -i
f^{abc}\hat J^c_R(l)\delta_{l,l^\prime}  \labe{comEU}\\
 && [\hat J^a_R(l), \hat J^b_L(l^\prime)] = 0 \qquad ,\nonumber
\end{eqnarray}
where $f^{abc}$ is the structure constant of the group and we choose the
normalization $2Tr\{T^aT^b\}=\delta^{a,b}$. $\hat{J}^a_R(l)$ and
$\hat{J}^a_L(l)$ are related using the adjoint representation $D^{(1)}_{ab}$
\begin{equation}\labe{JABD}
  \hat{J}^a_L(l) =   D^{(1)}_{ba}\left( U(l) \right) \hat{J}^b_R(l)
 = 2Tr\left[ U(l) T^b U^\dagger(l) T^a \right]\hat{J}^b_R(l) \qquad.
\end{equation}
The standard Kogut-Susskind lattice hamiltonian \cite{Kogut},
which we shall use, reduces  in two dimensions and in the absence of
dynamical fermions to the sum of the quadratic casimir operators
\begin{eqnarray}
   \hat{H}= \frac{g_L^2}{2a}\sum_{link}\hat{J}_R^a(l)\hat{J}_R^a(l)
 = \frac{g_L^2}{2a}\sum_{link}\hat{J}_L^a(l)\hat{J}_L^a(l)  ,\labe{sH}
\end{eqnarray}
where $g_L$ is the dimensionless lattice coupling constant and a is the lattice
spacing. The hamiltonian is invariant under
time independent gauge transformations generated by Gauss law operators
$\hat{G}^a(l)$
\begin{equation}
 \hat{G}^a(l)
 = \{ \hat{J}^a_L(l) - \hat{J}^a_R(l+1)
                + \psi^{\dagger} (l) T^a \psi (l) \} ,\labe{Sgauss}
\end{equation}
through which static color sources enter the formulation.
Since physical states have to be gauge invariant, we have to impose
the constraints
\begin{equation}\labe{consphy}
  \hat{G}^a(l) | \enskip phys > = 0 .
\end{equation}
In the following we consider only the case with one static quark and one static
anti-quark, which will also provide the solution in the case without
static sources in the limit of zero spatial separation.

\section{The spectrum of states}\labe{SSS}
\mbox{\indent}
We choose the sites $0$ and $m$ to be occupied with the heavy quark and the
heavy anti-quark respectively. In order to find the spectrum of gauge invariant
states we introduce the products $A$ and $B$ of matrices $U(l)$
\begin{eqnarray}
       & &   A = U(m) \times \cdots \times U(1)  \nonumber  \\
       & &   B = U(M) \times \cdots \times U(m+1)  .\labe{defAB}
\end{eqnarray}
These particular combinations are useful
since it is obvious that by combining $A$, $B$ and static sources gauge
invariant states can be obtained. For example when acting on the gauge
invariant ground state $|0>$ of eq.(\ref{sH}) with the property
\begin{equation}
\hat{J}_R^a(l)| 0 > = 0  \qquad \left(\quad\hat{J}_L^a(l)| 0 > = 0
                 \quad \right) \qquad\labe{defvac}
\end{equation}
gauge invariant states  are created from the operators
$\psi^{\dagger}\hat{A} \psi$, $\psi^{\dagger}\hat{B}^\dagger \psi$,
$\psi^{\dagger}\hat{A}\hat{B}\hat{A} \psi$, $Tr[\hat{A}\hat{B}]
\psi^{\dagger}\hat{A} \psi$ etc..
In order to be able to treat
( $ U(1), \ldots, U(m-1), U(m+1), \ldots, U(M-1), A, B $ ) as independent
variables, the angular momentum operators must be transformed accordingly.
Decomposing $A,B$ in terms of particular link variables $U(l)$ and auxiliary
matrices $L(l),R(l)$
\begin{equation}
  A = L_A(l)U(l)R_A(l) \ ,\qquad   B = L_B(l)U(l)R_B(l)\  ,
\end{equation}
the operators $J^a_R(l)$ are modified in the following way
\begin{eqnarray}
\begin{array}{ll}
  l \in [1,m-1]
 & J^a_R(l) \longmapsto  J^a_R(l) + D^{(1)}_{ba}(R_A(l))J^b_R(A)  \\
  l = m
 & J^a_R(l) \longmapsto  D^{(1)}_{ba}(R_A(l))J^b_R(A)    \\
  l \in [m+1,M-1]
  & J^a_R(l) \longmapsto  J^a_R(l) + D^{(1)}_{ba}(R_B(l))J^b_R(B)   \\
 l = M
 & J^a_R(l) \longmapsto  D^{(1)}_{ba}(R_B(l))J^b_R(B) .
\end{array}\labe{SchJ2}
\end{eqnarray}
In the new set of variables the constraints (\ref{Sgauss}) now read
\begin{eqnarray}
   & &(\enskip l \neq 0,m \enskip)
   \quad \left\{
   \begin{array}{c}
       J^a_L(l) | \enskip phys > = 0      \\
       J^a_R(l) | \enskip phys > = 0      \\
   \end{array} \quad \right. \labe{SnconJ1}  \\
&&\{ J^a_L(A) - J^a_R(B) +\psi^{\dagger} (m) T^a \psi (m)\} |\enskip phys > = 0
             \labe{snconj2} \\
&&\{ J^a_L(B) - J^a_R(A) +\psi^{\dagger} (0) T^a \psi (0)\} |\enskip phys > = 0
     \qquad \labe{SnconJ2}
\end{eqnarray}
which tell us that only the variables $A$ and $B$, as expected, are relevant
to construct physical states.
Owing to the replacement (\ref{SchJ2}), the hamiltonian takes the form
\begin{equation}
         \hat{H} = \frac{g_L^2d}{2a^2}\hat{J}^a_R(A)\hat{J}^a_R(A)
          +\frac{g_L^2(L-d)}{2a^2}\hat{J}^a_R(B)\hat{J}^a_R(B)+Q ,\labe{SnH}
\end{equation}
where $d=ma$ ($L=Ma$) is the distance between the two sources and the
operator $Q$ has the properties
\begin{equation}\labe{wSQ}
   Q | phys > = 0;\qquad \left[Q,J_R^a(A)J_R^a(A)\right] =
   \left[Q,J_R^a(B)J_R^a(B)  \right]=0\ .
\end{equation}
The constraints, which remain to be implemented, are given in
eq.(\ref{snconj2},\ref{SnconJ2}).
The problem of finding the spectrum of physical states has therefore been
reduced to that of two "link" variables of length $d$ and $L-d$ respectively.
Due to the one dimensional geometry the local degrees
of freedom have disappeared by imposing gauge invariance. \par
To find the eigenvalues and the eigenfunctions of the hamiltonian (\ref{SnH})
we restrict our consideration for simplicity to SU(2). The hamiltonian is a sum
of quadratic casimir operators which mutually commute. Therefore, in SU(2),
the eigenfunctions have the form of a direct product ($j_A $, $j_B$ ), labeled
by an A-spin, $j_A $, and a B-spin, $j_B$, respectively. Owing to the
constraints
(\ref{snconj2},\ref{SnconJ2}), these two representations, however,
have to fulfill the requirement $\vert j_A - j_B \vert = \frac12$.
As an expansion in terms of characters\footnote{There is another way of
formulating the eigenfunctions of the hamiltonian in terms of D--functions
which is less intuitive for our purposes but is easier to generalize to any
SU(N) group.}  $\hat{\chi}_{\frac{k}{2}}(BA)$ satisfying
\begin{eqnarray} \labe{chipro}
\hat{J}^a_R(U)\hat{J}^a_R(U) \enskip \hat{\chi}_j(U)| 0 > =
  j \left( j +1 \right)
     \hat{\chi}_j(U)| 0 >
\end{eqnarray}
we find the following expressions ($n=0,1,2,\dots$) \cite{Soh}
\begin{eqnarray}\labe{Snsol11}
   |\enskip (\frac{n}{2}+ \frac12 ,\frac{n}{2}) >_{phys}
     = \sum_{k=1}^n \hat{\chi}_{\frac{k}{2}}(\hat{B}\hat{A})
     \enskip[\hat{A}(\hat{B}\hat{A})^{(n-k)}]_{ij}
       \psi_i^{\dagger}(m) \psi(0)_j\enskip | 0 >
\end{eqnarray}
\begin{eqnarray}\labe{Snsol22}
   |\enskip (\frac{n}{2} ,\frac{n}{2}+ \frac12) >_{phys}
     = \sum_{k=1}^n \hat{\chi}_{\frac{k}{2}}(\hat{B}\hat{A})
    \enskip \left[\hat A[(\hat{B}\hat{A})^{(n-k+1)}]^{\dagger}\right]_{ij}
       \psi^{\dagger}(m)_i \psi_j(0)\enskip | 0 > \quad.
\end{eqnarray}
The eigenvalues corresponding to these states are
\begin{eqnarray}
  & & E(\frac{n}{2}+\frac12,\frac{n}{2})
   = \frac{g_L^2}{8a^2} \left[ d(n+1)(n+3) + (L-d)n(n+2)\right]
\labe{Segval1}\\
  & & E(\frac{n}{2},\frac{n}{2}+\frac12)
   = \frac{g_L^2}{8a^2} \left[ (L-d)(n+1)(n+3) + d\cdot n(n+2)\right]
                           \labe{Segval2} \  .
\end{eqnarray}
Since in this model the continuum limit can be taken trivially
($g_L/a\rightarrow g_{continuum}$) we find for SU(2) identical results as
obtained in
\cite{Burkardt} and which have also been found in the standard continuum
formulation \cite{Schreiber}.
\par
This agreement also holds in the case of pure Yang-Mills theory which is
obtained from the above result in the limit $d\rightarrow 0$
($A_{ij}\rightarrow
\delta_{ij}$). We find the following spectrum and eigenstates
(the states (\ref{Snsol22}) still couple to the sources in this limit
and therefore they do not describe pure gauge theory)
\begin{equation}
|\enskip \frac{n}{2} >_{phys}  = \hat{\chi}_{\frac{n}{2}}(\hat B\hat A)|
\enskip  0 >\  ,\quad  E(\frac{n}{2})
   = \frac{g_L^2L}{2a^2} \frac{n}{2} \frac{n+2}{2}
\end{equation}
given simply by the characters $\hat{\chi}_{\frac{n}{2}}(\hat B\hat A)$
of SU(2) and agreeing with results obtained in \cite{Rajeev}.


\section{Comparison with the continuum formulation}\labe{FPD}
In this section we intend to compare our lattice results with those of the
continuum formulation not only for the spectrum but also for the states
themselves. This enables us to interpret the zero mode variables in the
continuum approach and to draw conclusions about the importance of
related "lower dimensional fields" in higher
dimensions. We start by observing that the states (\ref{Snsol11},\ref{Snsol22})
are characterized only by the combinations of $\hat B\hat A$. We therefore
perform a transformation from $\hat A,\hat B$ to the variables $\hat A,
\hat U=\hat B\hat A$ in analogy to the transformation we carried out in the
previous section. The hamiltonian and the constraints for arbitrary
SU(N) groups then take the form
\begin{eqnarray}\labe{SnH2}
   \hat{H} &= &\frac{g_L^2d}{2a^2}\left\{\hat{J}^a_R(A)
           \hat{J}^a_R(A)+2\hat{J}^a_R(A)\hat{J}^a_R(U)
           \right\} +\frac{g_L^2L}{2a^2}\hat{J}^a_R(U)\hat{J}^a_R(U) \\
0 &=& \{ J^a_L(U) - J^a_R(U) -J^a_R(A)
         +\psi^{\dagger} (0) T^a \psi (0)\} |\enskip phys > \labe{NABGaussU}
\\
0 &=&\{ J^a_L(A)  +\psi^{\dagger} (m) T^a \psi (m)\} |\enskip phys >
             \qquad.  \labe{nabgaussU}
\end{eqnarray}
For comparison with the continuum formulation we next transform to the
axial gauge.  On the lattice this means we eliminate $\hat A$ and we
diagonalize $\hat U$, which can be achieved by applying two unitary
transformations $K_1$ and $K_2$
\begin{eqnarray}
  K_1 &=&\exp(-i\rho^a_m \omega^a_A) \ , \labe{defT1}\quad
        \rho^a_m = \psi^{\dagger} (m) T^a \psi (m) \\
  K_2 &=&\exp(-iq^a \Delta^a)\labe{defT2}\ ,\quad
   q^a = \psi^{\dagger} (m) T^a \psi (m)+
                    \psi^{\dagger} (0) T^a \psi (0)\ ,
\end{eqnarray}
where the angles $\omega^a_A$ and $\Delta^a$ are related to $A$ and $U$ as
\begin{equation}
A=\exp(i\omega^a_AT^a) ,\qquad
U = \exp (i\Delta )\exp (i\theta)\exp (-i\Delta ),  \labe{parU}
\end{equation}
with $\exp (i\theta)$ diagonal. For later use we introduce auxiliary
matrices $R$ and $P$ by
\begin{eqnarray}
R_{ab} &=& 2Tr\left[\exp (-i\Delta )T^a\exp (i\Delta )T^b \right] \labe{defD}\\
P_{ab} &=&
2Tr\left[\exp (-i\theta)T^a\exp (i\theta)T^b\right]-\delta_{ab}\labe{defP}\ .
\end{eqnarray}
Transforming the constraints (\ref{NABGaussU},\ref{nabgaussU})
we find\footnote{Note that we use indices with subindex 0 to enumerate the
elements belonging to the Cartan subalgebra and indices  with subindex 1
to enumerate the remaining group elements.}
\begin{eqnarray}
J^a_R(A) |\enskip phys >_K = 0  ,\quad
R_{ba_1}J^b_R(U)|\enskip phys >_K = 0 ,\quad
q^{a_0}|\enskip phys >_K = 0 \labe{Fcons2},
\end{eqnarray}
where  the subscript $K$ on physical states signals that these states differ
from the original ones by application of the unitary transformation. Using
these constraints  the unitarily
transformed hamiltonian $ \hat{H}'=K_2K_1\hat{H}K_1^{\dagger}K_2^{\dagger} $
reads in the physical Hilbert space
\begin{eqnarray}
  && \hat{H}^\prime=\frac{g_L^2d}{2a^2}\left\{\rho^a_m\rho^a_m
-2\rho^{a_0}_m\frac{i\partial}{ \partial \theta^{a_0}}
         + 2\rho^{a_1}_mP^{-1}_{a_1b_1}q^{b_1}
           \right\} +\frac{g_L^2L}{2a^2}
   \left\{P^{-1}_{a_1b_1}q^{b_1}P^{-1}_{a_1c_1}q^{c_1}\right.  \nonumber\\
&& - \left.\frac{\partial ^2}{\partial \theta^{a_0}\partial \theta^{a_0}}
          -\sum_{n<m=1}^N\cot  \left(
   \frac{T^{a_0}_{nn}-T^{a_0}_{mm}}{2}\theta^{a_0} \right)\left(T^{a_0}_{nn}-
     T^{a_0}_{mm}\right)\frac{\partial}{\partial \theta^{a_0} }\right\}
\labe{fsh} \quad .
\end{eqnarray}
The hamiltonian describes the dynamics of $N-1$
physical variables $\theta^{a_0}$ in the presence of the
constraints on the fermionic charges eq.(\ref{Fcons2}). According to
these constraints we have to consider $N$ basis states, one "singlet" state
and $(N-1)$ states in the adjoint representation.

In SU(2) there is only one variable ($\theta^3$) and the states can
be classified as "singlet" $|S>$ and "triplet" $|T>$ from which
general physical states are formed by linear superposition
\begin{eqnarray}\labe{fbase1}
|S>&=&\frac{1}{\sqrt{2}}\psi^{\dagger} (m)  \psi (0)|0>,\qquad
|T>=\frac{1}{\sqrt{2}}\psi^{\dagger} (m) 2T^3 \psi (0)|0>\quad , \\
|phys>_K &=& {\cal A}_S(\theta^3)|S>+{\cal A}_T(\theta^3)|T>.
\end{eqnarray}
The hamiltonian (\ref{fsh}) written in two-component form then reads
\begin{equation}\labe{SDH}
  H_{eff} =  -\frac{g_L^2L}{2a^2} \left[{{\partial ^2} \over {\partial
 {\theta^3}^2}}
   +\cot (\theta^3/2){\partial \over {\partial \theta^3}}\right]  +
  \frac{g_L^2}{8a^2}\left\{ {\matrix{{3d}&{-4id\cot (\theta^3/2)
  - 4id{\partial/{\partial \theta^3}}}\cr
        {- 4id{\partial/{\partial \theta^3}}}&{-d+2L
        \sin ^{-2}(\theta^3/2 )}\cr }} \right\}\ .
\end{equation}
The unusual appearance of  the differential operators in the off-diagonal
elements
can be corrected for by a third unitary transformation $K_3$ of the form
\begin{equation}
  K_3=\exp(i\rho^3_m \theta^3\frac{d}{L}) . \labe{defT3}
\end{equation}
Applying this operator and rescaling $\theta^3=2\pi c$ we obtain the continuum
hamiltonian as well as the basis states which were used in \cite{Schreiber}
to determine the spectrum of states
\begin{eqnarray*}
  \tilde{H}_{eff} &=&  -\frac{g_L^2}{8a^2}\left[\frac{L}{\pi^2}{{{\partial ^2}
  \over {\partial c^2}}
   +2\pi\cot (\pi c){\partial \over {\partial c}}} +\frac{d^2}{L}
  -\left\{ \matrix{{3d}&{-2id\cot ( \pi c) }\cr
        {2id\cot (\pi c) }&{-d+ 2L
        \sin ^{-2}(\pi c)} \cr } \right\} \right]\\
  |\tilde{S}>&=&\frac{1}{\sqrt{2}}\psi^{\dagger} (m)
                 \exp(2iT^3 \pi c\frac{d}{L}) \psi (0)|0>,\
|\tilde{T}>=\frac{1}{\sqrt{2}}\psi^{\dagger} (m) \exp(2iT^3 \pi c\frac{d}{L})
                 2T^3 \psi (0)|0>.
\end{eqnarray*}
Thus we have not only shown that identical results are found in continuum and
lattice approach  but also obtained the means to translate the results
expressed
in one language into the other. For a general SU(N) group this allows
us to identify the variables $\theta^{a_0}$ (for SU(2) the variable c)
according to eq.(\ref{parU}) as a parametrization of the trace of the
Wilson loop $U$ winding around the circle in 1+1 dimensions. In higher
dimensions these zero modes become "lower dimensional fields" maintaining
the coordinate dependence in the orthogonal directions.

\section{Discussion and conclusion}
For a discussion of the infrared properties of QCD$_{1+1}$ we observe that the
hamiltonians eqs.(\ref{fsh},\ref{SDH}) have characteristic singularities
in the variables $\theta^{a_0}$ similar to centrifugal barriers. The
"singlet" state eq.(\ref{fbase1}), which corresponds to the state n=0 in
eq.(\ref{Snsol11}), is the only eigenstate of the hamiltonian which has
no $\theta^3$ dependence and no L-dependent energy. Due to the singularities
in the hamiltonian, the $\theta^3$ dynamics is responsible for the L-dependent
level splitting between the "singlet" and the remaining states. It is, however,
not the origin of the d-dependent energy of the "singlet" state which can
be traced back to the operator $\rho_m^a\rho_m^a$ in eq.(\ref{fsh}) which
would be found in QED, as well.

In 2+1 or 3+1 dimensions the axial gauge hamiltonian shows singularities in
complete analogy to the 1+1 dimensional hamiltonian (\ref{fsh}). The difference
being the replacement of the zero modes $\theta^{a_0}$ by "lower dimensional
fields $\tilde\theta^{a_0}$ which depend on one or two spatial variables
respectively. This analogy, however, is not sufficient for arguing that
the "lower dimensional fields" will be as crucial as they are in 1+1
dimensions to understand the infrared
properties and in particular confinement of QCD in 2+1 or 3+1 dimensions.
The reason can be understood, if we remember that an excitation of the
$\theta^3$ variable is the same as the presence of the Wilson loop $U$, which
winds around the circle, in the lattice wave function. The coupling to this
particular Wilson loop is a necessity in 1+1 dimensions, if the fermions at
sites (0,d) do not couple to the gauge field string\footnote{Here we make use
of the continuum notation where P denotes path ordering}
$P\exp[ig\int_0^ddx\, A(x)]$ in a way to form a singlet with respect
to fermionic and gluonic color charge. In 2+1 or 3+1 dimensions
our comparison with the lattice calculation explicitely reveals that it will
be energetically favorable to couple to Wilson loops with
an extension in directions orthogonal to the zero mode direction of the
lower dimensional fields. In this way color singlet states may be formed
without causing L-dependent energies. In QED these "transverse" degrees of
freedom may be explicitly shown to change the linear potential to a Coulomb
potential. Of course this does not exclude that the
"lower dimensional fields" are important to understand the infrared
properties of QCD. It tells, however, that the similarity in the
singularity structure of the hamiltonian is not a guarantee for that.

Finally we want to mention a seemingly technical point. We observe that the
calculation of the spectrum of states in the axial gauge hamiltonian
eq.(\ref{fsh}), in which all redundant variables are eliminated, is in general
a very difficult task. In contrast, the lattice result
eqs.(\ref{Snsol11},\ref{Snsol22}) is easily generalized to any SU(N) group.
We believe that this is due to keeping unphysical degrees of freedom in the
lattice formulation. It should therefore not be an artifact of the 1+1
dimensional model but rather point out the possible disadvantage of
eliminating all unphysical degrees of freedom due to the complicated structure
of the resulting Hamiltonian.
{\Large\bf Acknowledgements}\\
This work has been carried out in part within the Monbusho International
Scientific Exchange Program "Quantum chromodynamics in a finite box" (No.
06044053). One of the authors (D.S.) gratefully acknowledges the financial
support by the "Japan Society for the Promotion of Science".

\end{document}